\documentclass[11pt,twoside]{article}
\usepackage{asp2010}
\usepackage{hyperref}

\resetcounters

\begin{document}

%\title{The know-how of differential phase}

\title{Circumstellar matter studied by spectrally-resolved interferometry}

\author{Florentin~Millour$^1$%, Romain~Petrov,$^1$. Philippe~Berio$^1$, and Anthony~Meilland$^1$R. Petrov, and A. Meilland, ajouter Denis et Olivier ?
\affil{$^1$ Lagrange lab., UMR7293, Univ. Nice, CNRS, OCA, Bd de l'Observatoire, 06304, Nice}}

\begin{abstract}
This paper describes some generalities about spectro-interferometry and the role it has played in the last decade for the better understanding of circumstellar matter. I provide a small history of the technique and its origins, and recall the basics of differential phase and its central role for the recent discoveries. I finally provide a small set of simple interpretations of differential phases for specific astrophysical cases, and intend to provide a "cookbook" for the other cases.
\end{abstract}

\section{Matter with movement which matters}

When asked to present "circumstellar matter studied with interferometry", I was put in a difficult situation, given the enormous amount of literature on such a subject \citep[see e.g. figure 2 in][]{LeBouquin2009}. In the 80's (when I was born) this would have been much easier than today, as only a handful of stars had been observed with the very few interferometers existing. In-between, large collecting area interferometers and general-user instruments have appeared, making the former observing achievements almost as easy as getting a spectrum with a spectrograph.

When speaking of matter around a central object, one can think of gas and/or dust, but one cannot separate its composition from its movement around that object. Indeed, a static sphere of dust would not wait very long until falling freely onto the centre of the system. Therefore, to make a stable system, one would need to impulse some movement to that sphere in order to counteract the gravity force:
\begin{itemize}
\item One would need, for example, to impulse rotation to this sphere in order to prevent it from free fall. But in this case, angular momentum effects would flatten it to a disk-like structure.
\item If expansion (or contraction) would be impulsed to the sphere, then nothing would change the shape of the sphere, but the gravity force would slow down (or accelerate) that sphere.
\item More complex physical effects could also occur:
\begin{itemize}
\item like in rotating and contracting viscous disks, where the loss of angular momentum would force the energy to escape from the system, either via heating, or via kinetic energy (materialized by jets).
\item such complex distribution of matter (disk+jet) would form a brake to an explosion of the central object (like in novae or supernovae), shaping the remnant nebula into bipolar outflows \citep{2007A&A...464...87W, 2007A&A...464..119C, 2011A&A...534L..11C}.
\item or like in turbulent flows, where the global structure of the flow (rotation/expansion) is respected, but not the local structure.
\end{itemize}
\end{itemize}

One can see here that nothing about the mass or the nature of the central object was hypothesized, and therefore, studying the kinematics of matter around a central object applies to many classes of objects with the same techniques. These objects range from the accretion disk around the white dwarf of a nova system, up to the accretion disk around the supermassive black hole in an active galactic nucleus (when sufficiently far away to neglect relativistic effects). Such accretion disks can also be found around a wide range of YSOs, from T Tauri stars \citep{2004Natur.432..479V} up to very massive stars \citep{Kraus2010}.

Studying such velocity fields has been made for decades with spectroscopy, but classical spectroscopy lacks the very important combination of velocity (spectroscopy) and position (astrometry) for all these compact objects. Interferometry provides this missing astrometric information, making possible that combination.
Therefore, I will try to restrain in this paper to the study of the combination of the distribution of matter and its movement around a central object.

\section{The origin of spectro-interferometry}

\subsection{At the beginning}

Speckle interferometry \citep{1970A&A.....6...85L} revolutionized high-angular resolution imaging in astronomy. However, in the first times of this technique, no phase measurement was attainable, making the obtained high-angular resolution information very difficult to interpret. Indeed, with only an access to the amplitude of the Fourier components of the image of the object, this technique was only able to provide autocorrelations of the object's image. One workaround of this limitation, called speckle holography, was introduced by \citet{1971BAAS....3..244G}, with a spectacular application to the resolution of the central object in R136a, in the core of the young stellar cluster 30~Dor \citep{1985A&A...150L..18W}. However, it needed an unresolved reference star inside the anisoplanetism angle of the atmosphere in order to work.

Until the advent of triple correlation techniques \citep{1993A&A...278..328H}, several ideas were proposed to recover the phases of the objects, by using their deterministic change as a function of spatial frequencies \citep{1974ApJ...193L..45K}, or as a function of wavelengths \citep{1979A&A....80L..13K}. However, the first to propose to compute phase differences between adjacent wavelengths in speckle interferometry as an observable is \citet{1983LowOB.167..165B}, with a direct potential application to close double stars research. In that work, the author proposed to use the properties of the cross-correlation of speckle patterns on the Multi-Mirror Telescope (MMT) to compute wavelength-dependent photocenter displacements of the telescope Point Spread Function (PSF). This displacement is linked directly to the nature of the object, here binary stars.
In \citet{1984A&A...134..354A}, the idea of Beckers was first tested on-sky. At this  time, the cross-correlation technique was used on single-dish telescopes (the MMT being considered here as single dish).

Later, this technique was applied to long-baseline interferometry on the \emph{Grand Interf\'erom\`etre \`a deux T\'elescopes} (GI2T), with some success \citep{1995A&A...300..219S, 1997A&A...323..183V, 1998A&A...335..261V, 1999A&A...345..203B}. Here, the cross-correlation technique was used to compute a phase as a function of wavelength on long-baseline interferometry data.

Spectroastrometry \citet{1998MNRAS.301..161B} was later invented, with the same purpose as the \citet{1983LowOB.167..165B} idea, i.e. study the photo-center shift as a function of wavelength, but with a different algorithm (in the image space rather than Fourier-space).

\subsection{Today's spectro-interferometry}

Today, spectro-interferometry has been popularized by the Very Large Telescope Interferometer (VLTI), with both AMBER \citep{Petrov2007} and MIDI \citep{1998AGM....14..L05G} instruments, joined by the Navy (Prototype) Optical Interferometer (NPOI, now NOI, the "prototype" aspect having disappeared now) in its differential phase referencing mode \citep{2009ApJ...691..984S}, VEGA/CHARA \citep{2009A&A...508.1073M}, and Keck used in self-referenced phase mode \citep{2012PASP..124...51W}.

The results coming from these instruments are significant, among which the direct detections of the formation zones of several ions, molecules, or dust types in young stellar objects \citep{2004Natur.432..479V, 2007A&A...464...43M, 2007A&A...464...55T, 2008A&A...489.1151T, Eisner2011}, the first direct detection of a Keplerian-rotating disk around a Be star \citep{2007A&A...464...59M} followed by many similar results on other Be stars \citep{2007A&A...464...73M, 2009A&A...504..915C, Stefl2012, 2012A&A...538A.110M},  the CO flow motion in the turbulent atmosphere of Betelgeuse \citep{Ohnaka2009, Ohnaka2011}, or the characterization of binary systems \citep{2007A&A...464..107M, 2008A&A...488L..67M, 2009ApJ...691..984S, Meilland2011a}.

More recently, a few attempts have been made to include the differential phases into an image-reconstruction process, either making assumptions on the target's resolution by the interferometer \citep{2009ApJ...691..984S}, or using radio-like image reconstruction methods in a more general case \citep{2011A&A...526A.107M, Ohnaka2011}. These recent attempts illustrate the fact that differential phase do add more information about the observed object than the one contained in classical interferometric observables alone, namely squared visibilities and closure phases, and therefore its huge potential for improving the image reconstructions with optical interferometers.

%This paper illustrates the philosophy behind the differential phase, the way one can compute it, a few typical examples of differential phases and the first-order interpretation that can be done with it.

\section{How to get it?}

The main idea behind computing interferometry observables is to get the \emph{invariant} part of the wandering fringes of an interferometer, as seen through the atmosphere and a spectrograph.

In the classical methods, this fringe wandering can be nulled in specific cases, where it has been demonstrated that no (or few) atmospheric effects affect the chosen observables.

For example, one can get the Fourier amplitudes using speckle techniques (i.e. using the power spectrum instead of directly the Fourier transform of the fringes). This calculation provides $V^2$ (squared FT amplitude) as a function of wavelength.

However, for computing the Fourier phases, things get more complicated:
\begin{itemize}
\item One way is to use the closure relation when using three telescopes, the very same way as for bispectrum speckle interferometry \citep{1993A&A...278..328H}. One phase (out of three) can therefore be extracted, independent from any instrumental or atmospheric effect. This phase is called "closure phase", but when using 3 telescopes, it misses $2/3$ of the available phase information.
\item Another way is to have at his disposal a model of the \emph{variable} part of these wandering fringes, that can be subtracted from the raw Fourier Transforms before averaging. One does not need to have a model of the \emph{observed object}, but rather a model of the \emph{atmosphere} and the \emph{instrument} to correct for these effects and calculate a phase (called "differential phase"). A by-product of this method is that one can get also the Fourier amplitude (called "linear visibility", or "differential visibility").
\end{itemize}

This last point is what is called \emph{differential} interferometry, because the instrument + atmosphere model is wavelength-dependent, and the observables are computed through a difference between the data and this model. The used model is time \emph{and} wavelength-dependent, and can be of growing complexity, depending on the wavelength coverage and spectral resolution of the instrument. The direct consequence is that this differential phase contains only a fraction of the original phase information.

This was illustrated in \citet[][pp. 58]{Millour2006}, \citet{Millour2008b}, and \citet{2009ApJ...691..984S}, where a Taylor-Expansion of the observed phase $\phi_{ij}^{m}(\sigma)$ as a function of wave-number $\sigma = {^1}/{_\lambda}$ was presented to illustrate the information kept into the differential phase, and the one discarded by the process. Indeed, the observed phase, in the absence of nanometer-accuracy metrology, is affected by an unknown optical path difference (OPD) term $\delta_{ij}(t)$, and higher order terms, which can come for example from the water-vapor and dry air chromatic dispersion $s_{ij}(t)$ :

\begin{equation}
  \phi^{m}_{ij}(t,\sigma) = \phi^{*}_{ij} (\sigma) + 2 \pi
    \delta_{ij}(t)  \sigma + 2 \pi
   s_{ij}(t)  \sigma^2 + ... + o(\sigma^n)
\label{eq:obsPhase}
\end{equation}

where $o(\sigma^n)$ is a negligible (or time-constant) term compared to all the other ones.
On the other side, one can perform a Taylor-Expansion of the object phase at the observation time:

\begin{equation}
  \phi_{ij}^{*}(\sigma)=a^{*}_{0}+a^{*}_{1}\sigma+a^{*}_{1}\sigma^2  + ... +
  \delta\varphi_{ij}^{*}(\sigma)
\label{eq:dvptPhase}
\end{equation}

where $ \delta\varphi_{ij}^{*}(\sigma)$ is the differential phase.
One can see here the similarities in the terms in in Eq.~\ref{eq:obsPhase} and \ref{eq:dvptPhase}. The differential data analysis process make disappear the 1-to-$n^{\rm th}$ first terms of this Taylor-Expansion. $n$ depends on the wavelength coverage and spectral resolution of the instrument. The different data processing processes are stacking the different contributions that can be modeled, depending on the spectral resolution. The higher the spectral resolution, the smaller will be $n$, because the smaller the variations of $\sigma$, making more and more of the higher-order terms negligible in the process.

\subsection{Getting the differential phases in practice}

\subsubsection{High spectral resolution}

This is the case of visible instruments (GI2T, VEGA), which have been designed and used with spectral resolutions of R=30,000, or the case of AMBER used in its highest spectral-resolution mode (R=12,000).

\begin{figure}[htbp] %  figure placement: here, top, bottom, or page
   \centering
\begin{tabular}{cc}
   \includegraphics[width=0.45\textwidth]{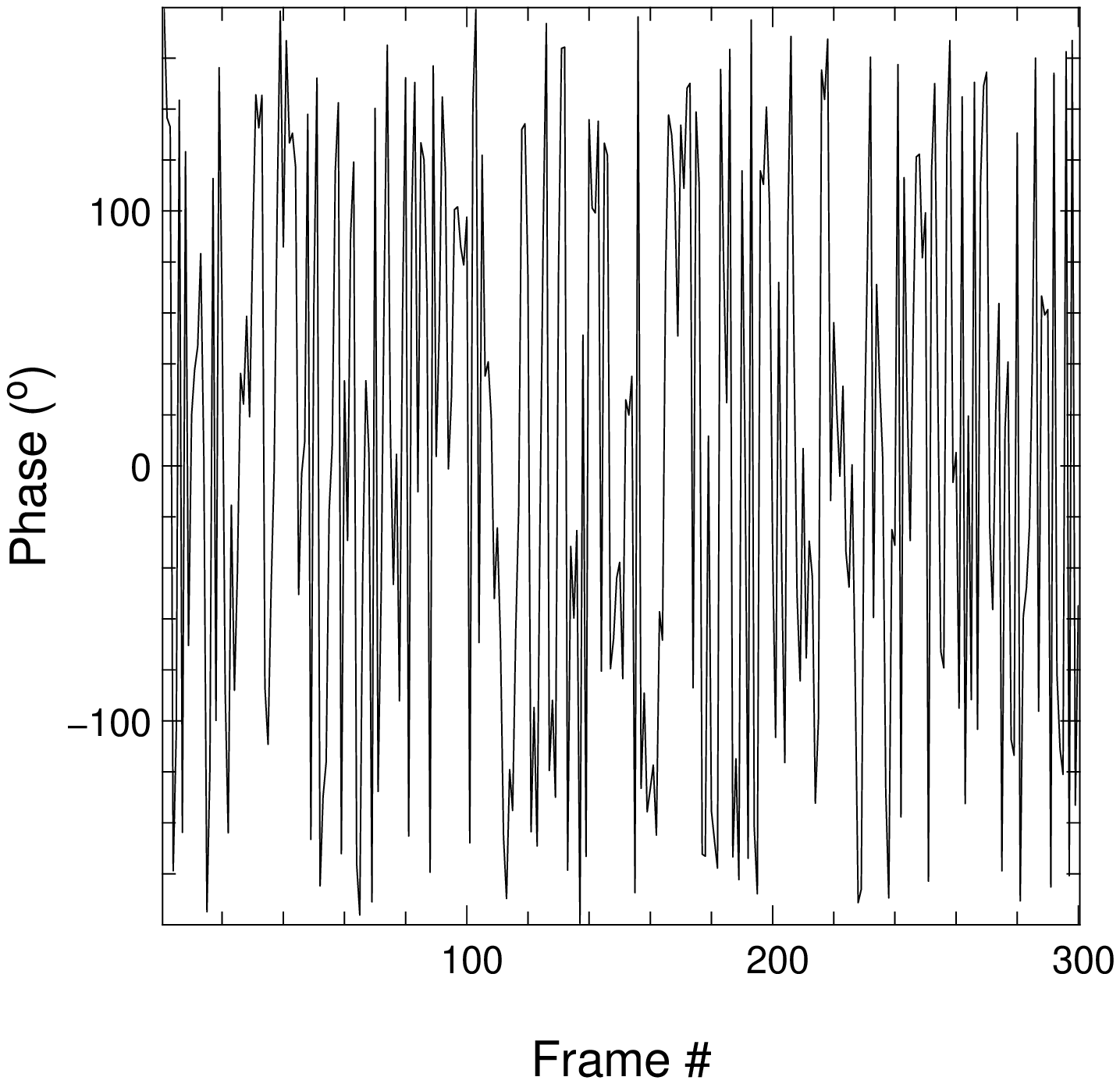}&
   \includegraphics[width=0.45\textwidth]{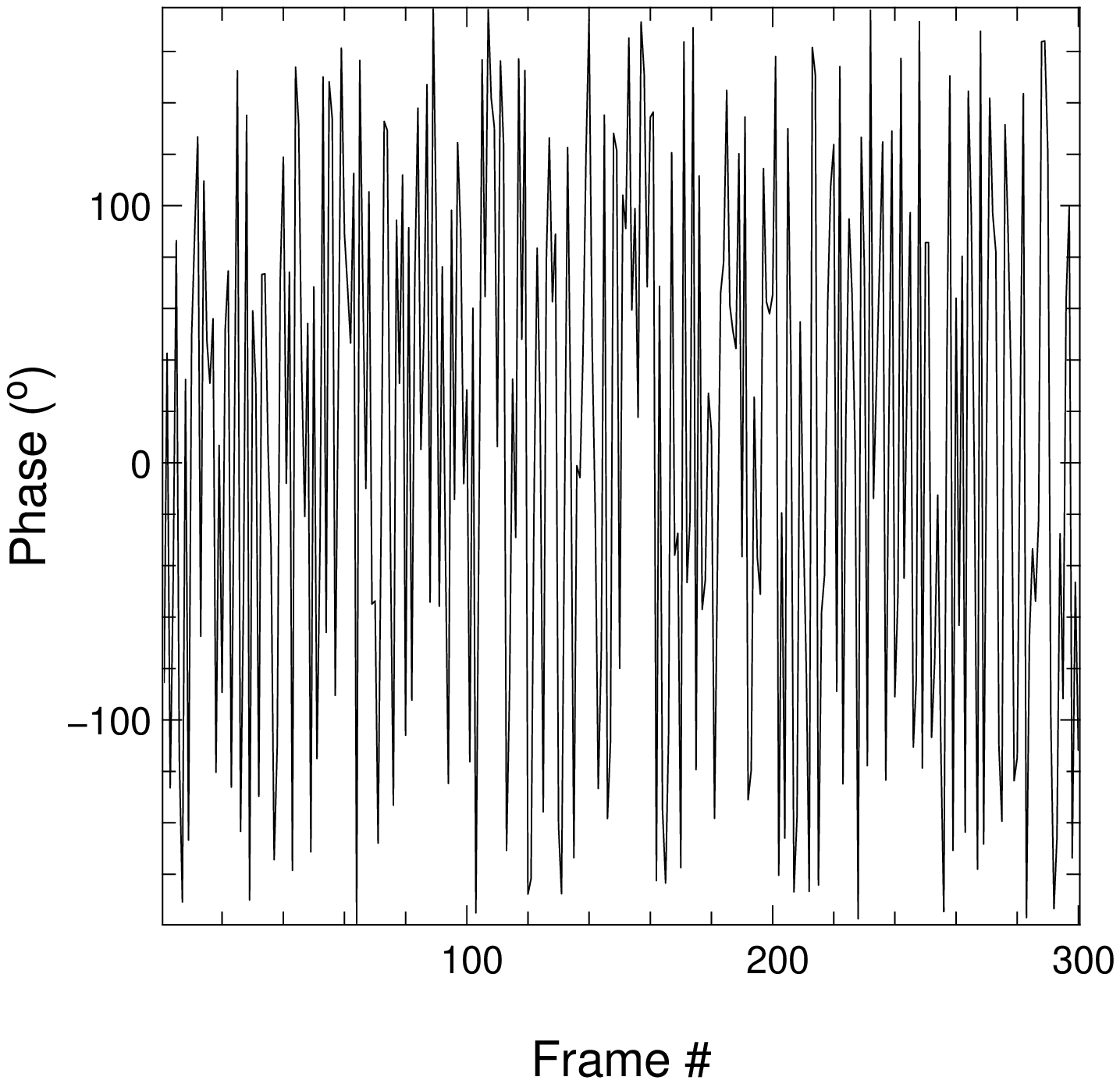}\\
    \end{tabular}
       \caption{Left: Phase of the AMBER fringes as a function of time, for a bright object (same as in Figure~\ref{fig:pistonWat}). Right: Phase of AMBER when no fringes at all are present.}
   \label{fig:PhaseInterf}
\end{figure}

In the case where the spectral interval of interest $\Delta\lambda$ is very small compared to the wavelength of interest $\lambda$, i.e. $\frac{\Delta\lambda}{\lambda} \ll 0.1$,
 the dominant atmospheric wandering of the fringes, due to the term $ 2 \pi \delta_{ij}(t)  \sigma$ of Eq.~\ref{eq:obsPhase}, can be considered as a constant as a function of $\sigma$. Therefore, that wandering is only an ensemble-movement of the fringes (see Figure~\ref{fig:PhaseInterf}), identical at all wavelengths. Therefore, the initial technique proposed by \citet{1983LowOB.167..165B} can be applied as-is, as its role is to remove the variable part of the phase.

Usually, the algorithms used to involve the computation of a cross-spectrum on each recorded frame, based on Fourier-Transforms or other algorithms (in the case of AMBER), and an ensemble averaging. At the very end of the processing, a $\sigma$-dependent residual slope of the phase may be subtracted, depending how good the white-light fringe was centered at observation time.

\subsubsection{Intermediate spectral resolution}

This is the case of several instruments such as AMBER, used in its medium-spectral resolution mode (R=1500), Keck-I in High-resolution mode (R=1000), or MIDI with high-spectral resolution (R=300).

\begin{figure}[htbp] %  figure placement: here, top, bottom, or page
   \centering
   \includegraphics[height=0.55\textwidth, angle=-0]{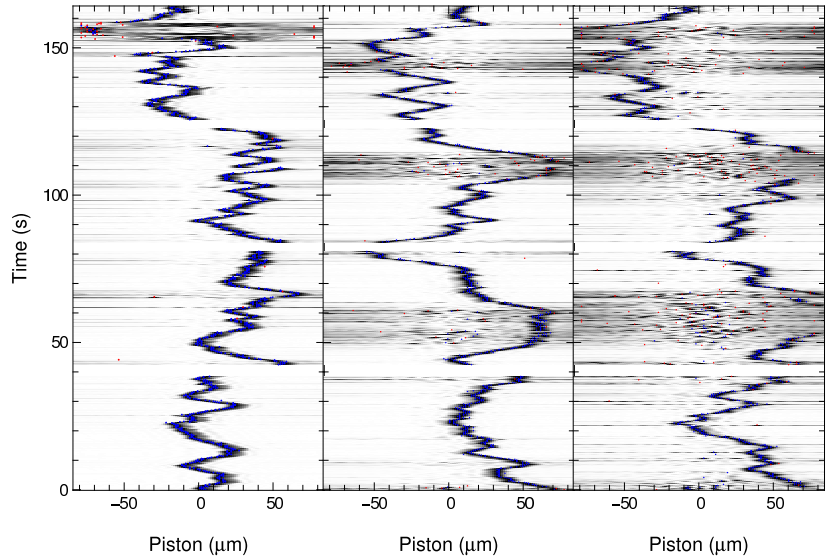}
          \caption{"Fringe-tracks" of AMBER/VLTI, showing the fringe wandering as a function of time (vertical axis). The plots show the fringe power as a function of optical path difference (piston, horizontal axis) for a bright star, shown here for the three baselines. The fringe wandering range is of the order of 10's of microns over a few seconds, with excursions as large as 50 to 100 microns.}
   \label{fig:pistonWat}
\end{figure}

In this intermediate case ($\frac{\Delta\lambda}{\lambda} \simeq 0.1$), more effects become non-negligible, like the $\sigma$-dependence of the term $2 \pi \delta_{ij}(t)  \sigma$, which imposes a calculation of $\delta_{ij}(t)$, and a frame-by-frame correction of this term (see Figure~\ref{fig:pistonWat}).

Therefore, for these instruments, a real-time correction (Keck, VLTI with FINITO), and/or a post-processing correction of this term are included in the observation/data processing pipeline \citep{2012PASP..124...51W, 2008A&A...489.1151T}.

\subsubsection{Low spectral resolution}

This is the case of NPOI (R=38), AMBER in its low-spectral resolution mode (R=35), or MIDI (R=30). In this case,  $\frac{\Delta\lambda}{\lambda} > 0.1$.

In addition to the two terms described above, additional terms, like the chromatic dependence of the OPD, or the variation of water vapor content, cannot be neglected anymore. Tackling and correcting for these effects is still today an active research topic \citep{2004SPIE.5491..588T, 2004PASP..116..876C, 2006MNRAS.367..825V, Millour2008b}.

\begin{figure}[htbp] %  figure placement: here, top, bottom, or page
   \centering
\begin{tabular}{cc}
   \includegraphics[width=0.45\textwidth]{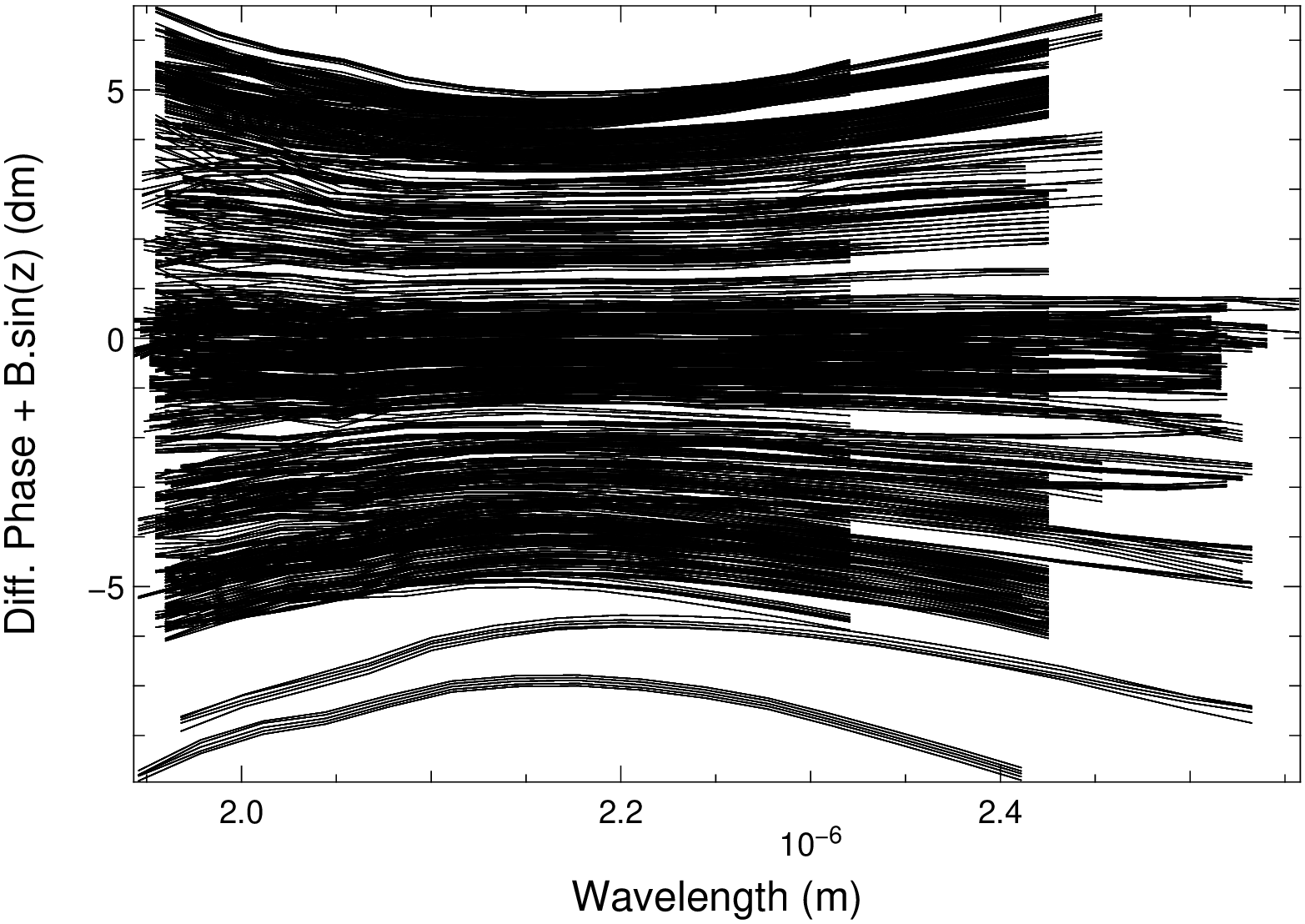}&
   \includegraphics[width=0.45\textwidth]{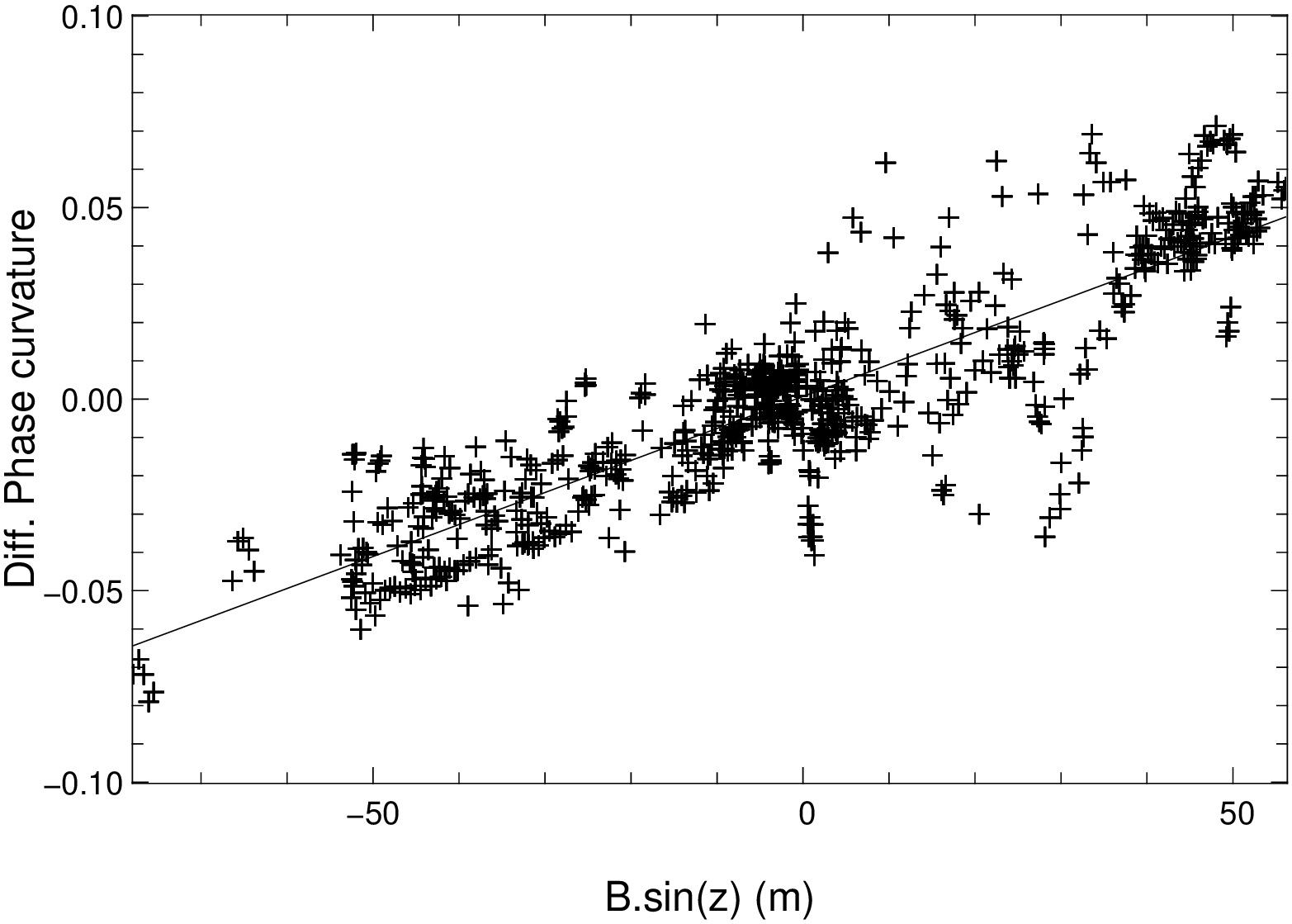}\\
   \includegraphics[width=0.45\textwidth]{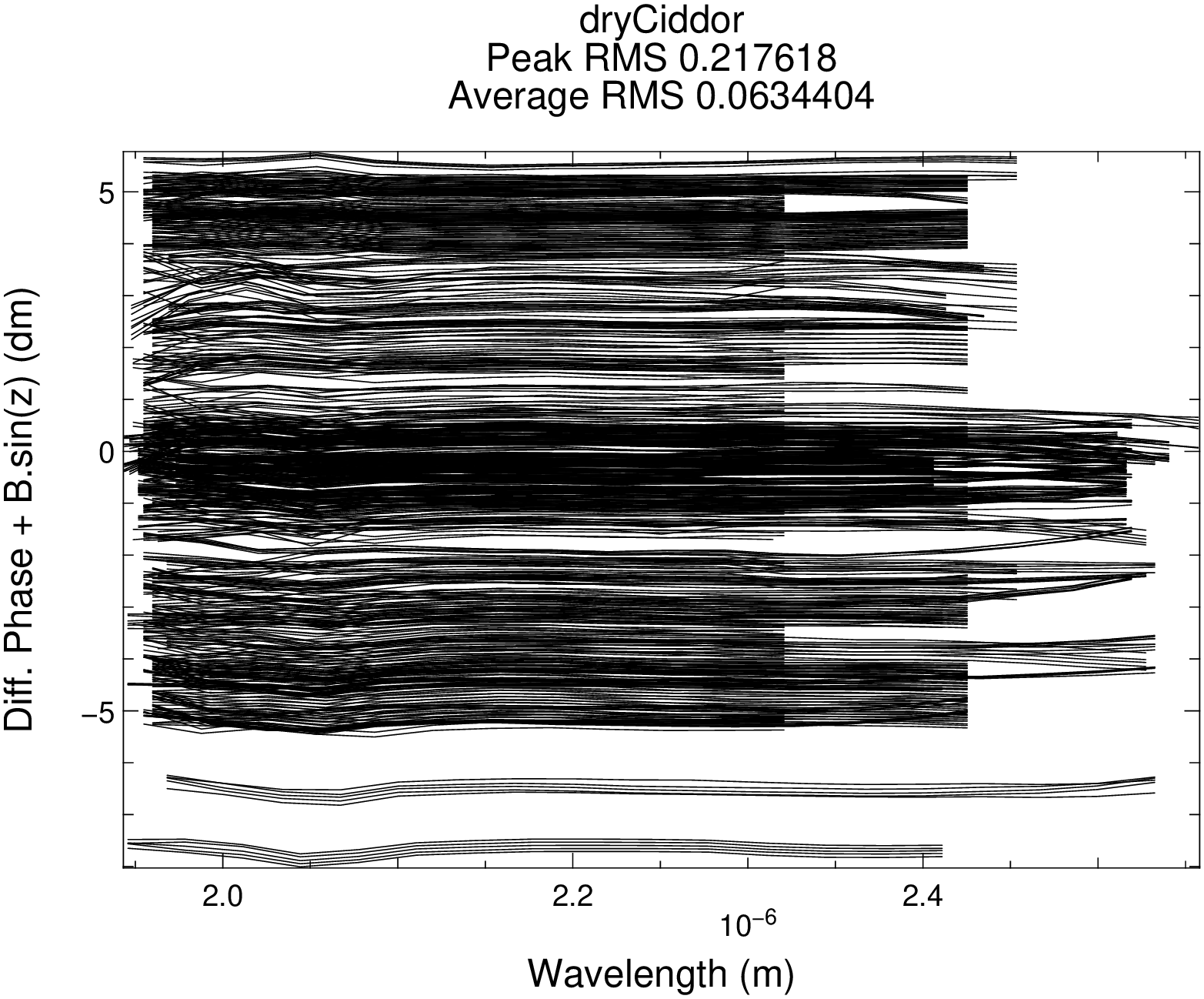}&
   \includegraphics[width=0.45\textwidth]{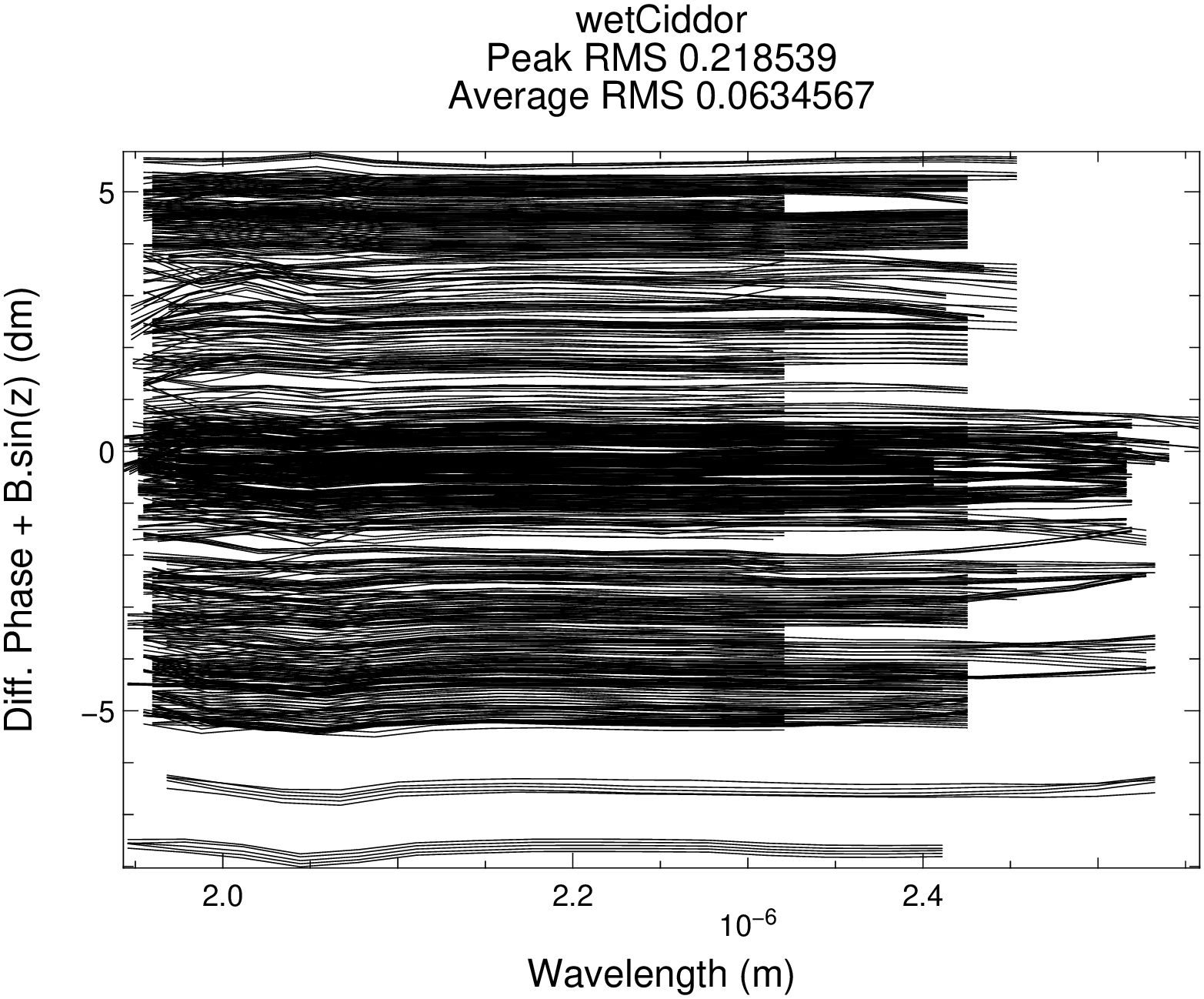}\\
    \end{tabular}
       \caption{From top to bottom and left to right: K-band raw differential phases from AMBER/VLTI with intrinsic noise less than 0.05 radians, for 34 full nights of AMBER operation (2008). They are offset by the $B\times sin(z)$ delay function (in decimeters) to exhibit the dependence of the phase curvature as a function of altitude. The following graph show the curvature ($2^{\rm nd}$ order derivative) of the differential phases as a function of delay. The corrected differential phases are shown afterwards, for dry air-only \citet{1996ApOpt..35.1566C}, and wet air \citet{1996ApOpt..35.1566C} corrections.}
   \label{fig:rawDataK}
\end{figure}

One can however, for an acceptable accuracy in many cases (2-3 degrees), tackle the slowly-variable part of this water-vapor + dry air term by pre-determined dispersion laws that can be found in the literature  \citep{1996ApOpt..35.1566C, 2006MNRAS.367..825V, 2007JOptA...9..470M}. Using the ambient conditions (temperature, pressure, humidity fraction, partial pressure of CO$_2$), measured simultaneously as the observation, and the delay lines position, one can compute the corresponding chromatic phase effect at the time of observing. An example of such correction can be found in Figure~\ref{fig:rawDataK}

\section{What does it mean?}

Differential phase is often thought to be either excessively simple, or too difficult to interpret. This matter of fact has several origins, which comes both from the poor understanding of the origin of that phase measurement (which is detailed in the previous section), and the fact that interpretation difficulty depends on how much the object is resolved by the interferometer. Indeed, as was shown in \citet{1995A&AS..109..389C, 2003A&A...400..795L, Vannier2006}, the relation between the differential phase and the photocenter of the object of interest is linear when the object is unresolved, or barely resolved (i.e. visibility is close to 1). Therefore, a direct comparison between spectro-interferometry and spectro-astrometry  can be done \emph{in the case the object is unresolved}.

One can then get a simple qualitative interpretation of their dataset, just based on its morphology. One knows already the "S"-shaped phases, that can be attributed to rotating disks \citep{1999A&A...345..203B}, but other possibilities appeared in the meantime, leading to a real "zoology" of differential signal:

\begin{itemize}
\item the classical "S"-shaped phase, associated to a "W" visibility \citep{2012A&A...538A.110M} in a line, is linked to a rotating disk,
\item on-sided phase, associated with a visibility drop in a line, can be associated to an asymmetric outflow or envelope \citet[see e.g.][]{Ohnaka2009, 2011A&A...535A..59B},
\item "double S" phase, or "W" phase, related to visibility variations in a line, can be interpreted with a bipolar outflow \citep[see examples in][]{2007A&A...464...87W, 2011A&A...534L..11C}.
\end{itemize} 

In all other cases, where the object is resolved by the interferometer, there is no simple relation between the phase and the photocenter shift, though in some cases (e.g. Be stars or B[e] stars), a "phase reversal" \citep[a change of phase variation, or a phase jump, see e.g.][]{2007A&A...464..119C, 2012A&A...538A.110M} can be attributed to a visibility lobe change.
Therefore, only a proper computation of the Fourier terms calculated on a model, or on an image, can help to interpret the datasets. The \emph{"cookbook of differential phase modeling"} could contain the following steps: 
\begin{itemize}
\item Get your freshly measured differential phase in a cool place, taking care to note its wavelengths and baselines,
\item Compute a pile of images of your model at as many wavelengths as your differential phase has been measured,
\item Fourier-Transform all your images, taking care of aliasing effects,
\item Interpolate carefully this FT at the exact spatial frequency of the observation, for each wavelength,
\item You should get a clean phase of your model for each wavelength and each baseline of observation,
\item Subtract as many terms as necessary from the obtained phase, using eq.~\ref{eq:dvptPhase},
\item The result is your modeled differential phase, that you can compare with your observed one.
\end{itemize}

\section{Conclusion}

In this proceeding, I have tried to show the tremendous advance made in the last years on circumstellar matter with the combination of interferometry and spectroscopy. This unique tool has not yet shown its huge potential in getting more physical information about the observed objects.

\section*{Questions}

\begin{enumerate}
\item[Q:]  (A. Magalhaes) Can you use the differential phase observations in order to model the observations, in the same way, for instance, that you do for polarization and position angle, where you compare with a model?
\item[A:] Yes, indeed, the differential phase can be computed based on the model images at several wavelengths, and compared to the observed one. This is actually the way we compare the models to the observations.
\item[Q:]  (P. Nu\~nez) Could you please explain the use of  "differential closure phase"?
\item[A:] "Differential closure phase" \& differential visibilities essentially provide the same information as the closure phase and squared visibilities, respectively. The main difference resides in the uncertainties and biases which affect these observables. For example, in specific cases, the "differential closure phase' can have significantly smaller errors than the closure phase.
\end{enumerate}

\bibliographystyle{asp2010}
\bibliography{biblio}

\end{document}